\documentclass[a4paper,11pt]{JHEP3}
\usepackage{graphicx,epsfig}
\pdfoutput=1

\def\be{\begin{equation}}
\def\EQ{\begin{equation}}
\def\ee{\end{equation}}
\def\EN{\end{equation}}
\def\bea{\begin{eqnarray}}
\def\beq{\begin{eqnarray}}
\def\ba{\begin{eqnarray}}
\def\eea{\end{eqnarray}}
\def\ena{\end{eqnarray}}
\def\eeq{\end{eqnarray}}
\def\ea{\end{eqnarray}}
\def\bei{\begin{itemize}}
\def\eei{\end{itemize}}
\def\bee{\begin{enumerate}}
\def\eee{\end{enumerate}}

\def\lx{\left}
\def\rx{\right}
\def\la{\langle}
\def\ra{\rangle}

\def\a{\alpha}
\def\b{\beta}

\def\lsim{\mathrel{\rlap{\lower3pt\hbox{\hskip0pt$\sim$}}
    \raise2pt\hbox{$<$}}}         
\def\gsim{\mathrel{\rlap{\lower2pt\hbox{\hskip1pt$\sim$}}
    \raise2pt\hbox{$>$}}}         
\def\sqr#1#2{{\vcenter{\vbox{\hrule height.#2pt
     \hbox{\vrule width.#2pt height#1pt \kern#1pt
           \vrule width.#2pt}
       \hrule height.#2pt}}}}

\def\noN{\nonumber}
\def\ccr{\nonumber\\}

\newcommand{\comment}[1]{}

\def\sideremark#1{\ifvmode\leavevmode\fi\vadjust{\vbox to0pt{\vss
 \hbox to 0pt{\hskip\hsize\hskip1em
 \vbox{\hsize3cm\tiny\raggedright\pretolerance10000
  \noindent #1\hfill}\hss}\vbox to8pt{\vfil}\vss}}}

\preprint{}

\title{Smooth tensionful higher-codimensional brane worlds with
  bulk and brane form fields}

\author{Olindo Corradini\\
Dipartimento di Fisica, Universit\`a di Bologna \\
and INFN, Sezione di Bologna \\ 
Via Irnerio 46, Bologna I-40126, Italy
\\ $\hskip3cm$ \& \\
Centro de Estudios en F\'isica y Matem\'aticas Basicas y Aplicadas\\
Universidad Aut\'onoma de Chiapas\\
Tuxtla Guti\'errez, Chiapas, Mexico\\ 
 
E-mail: \email{corradini@bo.infn.it}}

\abstract{
Completely regular tensionful codimension-$n$ brane
  world solutions are discussed, where the core of the brane is chosen
  to be a thin codimension-$(n-1)$ shell in an infinite volume flat
  bulk, and an Einstein-Hilbert term localized
  on the brane is included (Dvali-Gabadadze-Porrati models). In order
  to support such localized sources we enrich the vacuum structure of
  the brane by the inclusion of localized form fields. We find
  that phenomenological constraints on the size of the internal core
  seem to impose an upper bound to the brane tension. Finite
  transverse-volume smooth solutions are also discussed.}

\begin{document}

\section{Introduction and Summary}

Brane world models have drawn a lot of attention in the last years since
they provide an interesting scenario for the search of solutions 
to long standing particle physics puzzles
as the cosmological constant problem and the hierarchy problem. In
cosmology they might provide alternatives to dark matter and/or dark
energy (see e.g.~\cite{Koyama:2007rx,Dvali:2003rk}).  

In the present manuscript we study brane world models with codimension
larger than two, for a variety of situations. However, we are mostly
interested in flat bulk models where the extra-dimensional volume is
infinite and 4d gravity is brane-induced on the brane at short
scales~\cite{DGP,DG,Dvali:2002pe} (see~\cite{Kakushadze:2001bd} for an
orientifold derivation). 
Thin tensionful higher-codimensional solutions in flat space are
known to give rise to singular backgrounds~\cite{Gregory} and need to
be regulated. 
One possible way to regulate such singularity is to "resolve'' the
brane, by giving it a non-trivial core, in the
 extra-dimensions, e.g. a thin spherical
shell; in this latter case the brane results effectively codimension-one.  
This method has proven to be quite efficient in the
codimension-two case, both for finite volume rugby-ball~\cite{Carroll:2003db} solutions and
for infinite-volume induced gravity ones.\footnote{Tensionful
  codimension-two singularities are milder (conical) and do  
not, a priori, need to be regulated~\cite{KK1}. However,
regularization via smoothing out the brane profile
is often invoked in order to avoid subtleties
associated to purely conical radially symmetric extra-dimensional space~\cite{Cline:2003ak}.}

For codimension larger than two it had been shown that a 
naive regularization of the higher-codimension brane by blowing the
thin brane to a thin spherical shell lead to a no-go
theorem~\cite{Corradini:2002ta}, that we review 
later. A possible way-out to such no-go theorem was then found
in~\cite{Corradini:2002es} by employing bulk higher curvature
  terms to regulate the bulk singularity. Another recent proposal for
  smoothing out higher-codimensional singularities is to consider a
  bulk Einstein-Skyrme model~\cite{BlancoPillado:2008cp}.  
Here we present a different way-out: we keep
bulk Einstein-Hilbert gravity but consider  
a richer brane vacuum structure by the inclusion of
higher-rank (form) fields (this was suggested
in~\cite{Charmousis:2004zd} for a $Z2$-symmetric
setup~\cite{Ellwanger:2003fd}) similarly
  to the codimension-two models that involve an axion
  field~\cite{Aghababaie:2003,Peloso:2006cq,Papantonopoulos:2006dv,Kaloper:2007ap} 
whose vacuum expectation
  value cancels the tangential (to the brane profile) component of the pressure.  
 We explicitly
show here that the inclusion of higher-rank fields 
works as well for our higher-codimensional solutions. 

Another reason behind the present work is the study of new
(higher-codimensional) brane cosmology models: as said above
codimension-one regularization seems necessary for cosmological 
setups, at least in bulk Einstein gravity (for Lovelock gravity and/or
broken spherical symmetry the situation might
improve~\cite{Charmousis:2005ey,leetasinato,Corradini:2008tu}). 
Such regularization allowed to study some cosmological properties of codimension-two setups using
  the moving brane approach~\cite{Minamitsuji:2007fx} or
  weak field limit~\cite{Peloso:2006cq,Kobayashi:2007kv}. We can thus also see the present
  work  as a possible framework where study cosmology on a
  generic-codimension brane world. 

  Finally we consider higher-codimensional induced gravity brane world models, in the
  light of more recent results~\cite{deRham:2007xp} where it was found
  that cascading higher-codimensional induced-gravity models are
  ghost-free, hence shedding new light on such induced
  gravity models, which have been sources of several controversies
  regarding their classical and quantum stability. 
  In~\cite{deRham:2007rw} it was also suggested that
  cancellation of ghost excitations might as well take place for  
  resolved brane setups with codimension larger that two, provided
  tangential pressures are
  cancelled. We show later that, opposed to the codimension-two case,
  in our setup tangential pressures do not have to vanish and no strong
  fine-tuning between flux field and tension is a priori
  needed. 
However, phenomenological constraints on the
size of the internal brane profile seem to impose, for this class of
models, an upper bound (cfr. eq.~(\ref{tension-contraint})) to the
brane tension, as opposed to the codimension-two case
where the upper bound for the tension is due to a topological
constraint (the conical deficit angle is bounded to be less than $2\pi$). 


\section{Vanishing bulk cosmological constant}\label{section:flat}

{}The brane world model we study in this section is described 
by the following action:
\begin{eqnarray}\label{action}
 S&=&{\widehat  M}^{D-2} \int d^{D}x~
 \sqrt{-{ \widehat g}}~{\widehat{\cal R}}\noN\\&&+ \int_{\Sigma} d^{D-1}x~
 \sqrt{-{ g}}\Biggl[{M}^{D-3}\left({\cal R}-{\Lambda}-\frac{1}{2\cdot p!}
F_{[p]}^2\right)+2\widehat M^{D-2} K_\pm\Biggr]\\[2mm]
&&\Sigma = {\bf R}^{D-n-1,1}\times {\bf S}^{n-1}_\epsilon
\nonumber
\end{eqnarray}
Here  $\Sigma$ is a fat codimension-$n$  
source brane, whose geometry is given by 
the product ${\bf R}^{D-n-1,1}\times {\bf S}^{n-1}_\epsilon$, where 
${\bf R}^{D-n-1,1}$ is the $(D-n)$-dimensional Minkowski space, and 
${\bf S}^{n-1}_\epsilon$ is a $(n-1)$-sphere of radius $\epsilon$ (in the
following we will assume that $n\geq 3$).  
The quantity ${M}^{D-3} {\Lambda}$ plays the role of 
the tension of the brane $\Sigma$ and  
$F_{[p]}$ is the field strength of a $(p-1)$-form 
potential~$A_{[p-1]}$
\begin{equation}
F_{[p]} = d A_{[p-1]}~, \quad\quad F_{m_1\dots m_p} = p \partial_{[m_1} A_{m_2\dots m_p]} =
\partial_{m_1} A_{m_2\dots m_p} +{\rm cyclic}  
\end{equation}
Also,  
\begin{equation}
 {g}_{mn}\equiv
 {\delta_m}^M{\delta_n}^N \widehat g{}_{MN}\Big|_\Sigma~,
\end{equation} 
where $x^m$ are the $(D-1)$ coordinates along the brane (the $D$-dimensional
coordinates are given by $x^M=(x^m,r)$, where $r\geq 0$ is a non-compact
radial coordinate transverse to the brane, and the signature of the 
$D$-dimensional metric is $(-,+,\dots,+)$); finally, the $(D-1)$-dimensional
Ricci scalar ${\cal R}$ is constructed from the $(D-1)$-dimensional
metric $g_{mn}$ and $K$ is the extrinsic curvature, with $K_\pm
\equiv K_++K_-$. In the following we will use the notation
$x^i=(x^\alpha,r)$, where $x^\alpha$ are the $(n-1)$ angular 
coordinates on the sphere.
Moreover, the metric for the coordinates $x^i$ will be (conformally) flat:
\begin{equation}
 \delta_{ij}~dx^i dx^j=dr^2+r^2\gamma_{\alpha\beta}~d\theta^\alpha d\theta^\beta~,
\end{equation}
where $\gamma_{\alpha\beta}$ is the metric on a unit $(n-1)$-sphere.
Also, we will
denote the $(D-n)$ Minkowski coordinates on ${\bf R}^{D-n-1,1}$ via $x^\mu$
(note that $x^m=(x^\mu,\theta^\alpha)$).

{}The bulk equations of motion are clearly given by
\bea
\widehat G_{MN}=0
\eea
and the boundary conditions for the fat brane can be obtained using
Israel junction conditions
\bea
\Big\la K^m{}_n-\delta^m_n K\Big\ra_\pm = -\frac1{2 \widehat M^{D-2}}T^m{}_n
\label{israel}
\eea
where
\begin{eqnarray}
T_{mn} =-M^{D-3}\Big(2G_{mn}+g_{mn}\Lambda\Big) + T_{mn}(F) 
\end{eqnarray}
is the total energy-momentum tensor for the ``matter'' localized on the fat brane, 
with
\begin{equation}
T_{mn}(F) = \frac{M^{D-3}}{(p-1)!}\biggl( -\frac{1}{2 p} F^2 g_{mn} + 
F_m{}^{l_2\dots l_p} F_{n\, l_2 \dots l_p}\biggr)~.
\end{equation} 

\subsection{The no-go theorem}
\FIGURE
{
\epsfig{file=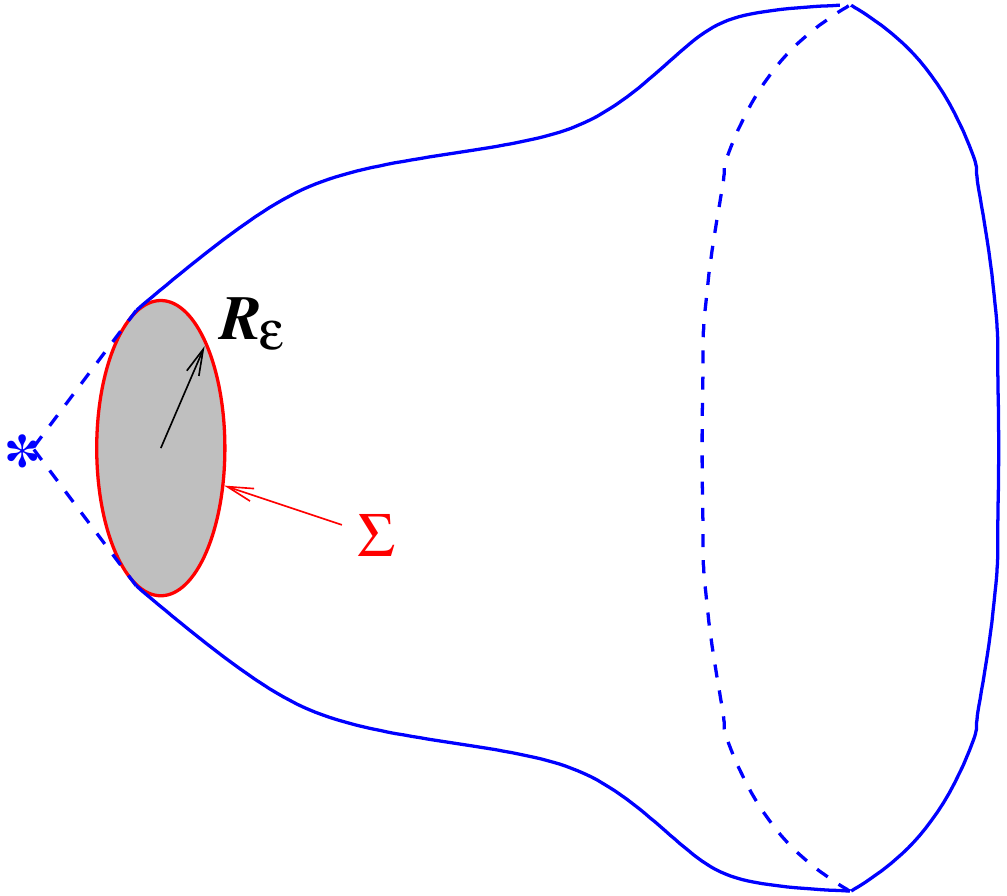,width=5cm}
\caption{Pictorial representation of the infinite-volume smooth brane
  world.}
\label{background_n}
}
In order to better clarify our results let us  first revise the no-go theorem
associated to radially symmetric solutions in absence of the $p$-form
term~\cite{Corradini:2002ta}. Let us consider the following ansatz for
the  background metric:
\begin{equation}
 ds^2=\exp(2A)~\eta_{\mu\nu}~dx^\mu dx^\nu+\exp(2B)~\delta_{ij}~dx^i dx^j~,
\end{equation}
where $A$ and $B$ are functions of $r$ but are independent of $x^\mu$
and $\theta^\alpha$ (that is, we are 
looking for solutions that are radially symmetric in the extra dimensions). 
The bulk equations of motion then read (here prime
denotes derivative w.r.t. $r$):
\begin{eqnarray}\label{AB1}
 &&(D-n)\left[\frac{D-n-1}{2}(A^\prime)^2+\frac{n-1}{r}A^\prime+
 (n-1)A^\prime B^\prime\right]
\ccr&&
+(n-1)(n-2)\left[{1\over 2}(B^\prime)^2+{1\over r}B^\prime\right]
 =0\quad \\[2mm]
 &&(D-n)\left[A^{\prime\prime}+
 \frac{D-n+1}{2}(A^\prime)^2+\frac{n-2}{r}A^\prime+(n-3)
 A^\prime B^\prime\right] \label{AB2}\ccr &&
+ (n-2)\left[B^{\prime\prime}+\frac{n-3}2(B^\prime)^2+\frac{n-2}{r}
 B^\prime\right]=0 
\end{eqnarray}
Above, equation~(\ref{AB2}) is the
$(\alpha\beta)$ equation, while equation~(\ref{AB1}) is the $(rr)$
equation.
Note that the latter equation 
does not contain second derivatives of $A$ and $B$.
The solution for $B^\prime$ is given by (we have chosen the plus root,
which corresponds to solutions with infinite-volume extra space):
\begin{equation}\label{root}
 B^\prime=-{1\over r}-{{D-n}\over{n-2}} A^\prime +
 \sqrt{{1\over r^2}+{1\over \kappa^2} (A^\prime)^2 }~,
\end{equation}
where we have introduced the notation
\begin{equation}
 {1\over \kappa^2}\equiv {(D-n)(D-2)\over (n-1)(n-2)^2}
\end{equation} 
to simplify the subsequent equations.


\comment{{}We can solve the above equations of motion as follows. First, we consider
the difference of (\ref{AB1}) and (\ref{AB2}) that, making use of
equation~(\ref{root}), reduces to 
\begin{equation}\label{Q}
 {Q Q^\prime\over\sqrt{1+Q^2}}+(n-2){1\over r}Q^2=0~,
\end{equation}
where 
\begin{equation}
 Q\equiv {1\over\kappa} rA^\prime~.
\end{equation}}
Here we are interested in non-singular
solutions such that $A$ and $B$ are constant
for $r<\epsilon$, and asymptote to some finite values as $r\rightarrow\infty$.
For $r>\epsilon $ the solution 
\comment{for $Q(r)$ is given by:
\begin{eqnarray}
 &&Q(r)={2f(r)\over {1-f^2(r)}}~,
\end{eqnarray}
where $f(r)\equiv \left({r_*\over r}\right)^{n-2}$, 
and $r_*$ is the integration constant. 
Next, we solve} for $A$ and $B$ is given by
\begin{eqnarray}
 &&A(r)=-{\kappa\over{n-2}}~
 \ln\left({{1+f(r)}\over{1-f(r)}}\right)~,~~~r>\epsilon~,\\
 &&B(r)=-{{D-n}\over{n-2}}A(r)+{1\over{n-2}}~
 \ln\left(1-f^2(r)\right)~,~~~r>\epsilon~,
\end{eqnarray}
where $f(r)\equiv \left({r_*\over r}\right)^{n-2}$ and $r_*$ is the
integration constant, and where we have set other integrations
constants such that $A_\infty = B_\infty=0$. 
A pictorial
representation of such setup is given in Fig.~\ref{background_n}, where the gray
disk describes the extra-dimensional shape of the inside
bulk ($r<\epsilon$), the bell-shaped part is the asymptotically-flat outside
bulk ($r>\epsilon$), the circle $\Sigma$ is the fat brane and the "star''
represents the would-be naked singularity $r=r_*$. 
Israel junction conditions~(\ref{israel}) provide the equations
at the location of the fat brane, $r=\epsilon$; including the
contribution of the induced EH term 
\bea
&& G^\mu{}_\nu = -\frac{(n-1)(n-2)}{2R_\epsilon^2}\delta^\mu_\nu\\
&& G^\a{}_\b = -\frac{(n-3)(n-2)}{2R_\epsilon^2}\delta^a_\b
\eea
where
$ R_\epsilon \equiv \epsilon\ e^{B(\epsilon)} $
is the physical radius of the $(n-1)-$sphere, we obtain
\bea\label{match2}
&& (n-2){2f^2(\epsilon)\over{1-f^2(\epsilon)}} +
 {\epsilon{  L}\over 2}
 e^{B(\epsilon)}\left[{ \Lambda}-\frac{(n-2)(n-3)}{R_\epsilon^2}\right]=0\\[2mm]
&&(n-1){2f^2(\epsilon)\over{1-f^2(\epsilon)}}-{{D-2}\over {n-2}}
 {2\kappa f(\epsilon)\over{1-f^2(\epsilon)}} +
 {\epsilon{  L}\over 2}
 e^{B(\epsilon)}\left[{
     \Lambda}-\frac{(n-1)(n-2)}{R_\epsilon^2}\right]=0
\label{match1}
\eea
for the $(\a\b)$ and $(\mu\nu)$ components respectively, with 
$L\equiv \frac{M^{D-3}}{\widehat M^{D-2}}$. 
We can rewrite the previous matching conditions in a more useful way 
as follows:
\begin{eqnarray}
 &&{2f^2(\epsilon)\over{1-f^2(\epsilon)}}+{{  L}\over 2 R_\epsilon}
 (\lambda-n+3)=0~,
\label{match1'}\\[2mm]
 &&\frac{D-2}{n-2} {2\kappa f(\epsilon)
 \over{1-f^2(\epsilon)}}+{{  L}\over 2 R_\epsilon}
 ( \lambda+n-1)=0~.
\label{match2'}
\end{eqnarray}
where we have defined
$ {  \Lambda}\equiv \lambda\frac{n-2}{R_\epsilon^2}~.$
Let us study possible solutions to these matching conditions with 
$r_*<\epsilon$ for which $0<f(\epsilon)<1$: they would be non-singular
solutions as the would-be naked singularity $r=r_*$ is cut away. The
second matching conditions can only be satisfied if ${ \Lambda}<0$. 
Hence $\lambda$ must be a negative parameter; in other words
there are no non-singular solutions of this type with positive tension. Moreover,
from the first condition we have:
\begin{equation}
 f(\epsilon)={{-\lambda+n-3}\over {-\lambda-n+1}}~\sqrt{(n-1)(D-2)\over{D-n}}~.
\label{f}
\end{equation}
For $n\geq 3$, the condition $0< f(\epsilon) < 1$
admits no solutions with negative $\lambda$. Hence, the above matching conditions cannot be
simultaneously satisfied within this class of solutions. For a
different class of solutions that is curved both on the inside bulk 
and on the outside bulk it is possible to overcome the previous no-go
theorem~\cite{Ellwanger:2003fd}. In~\cite{Charmousis:2004zd} an
upgraded version of the model~\cite{Ellwanger:2003fd}, that suggested
the use of brane form fields, was considered. 
 In the next section we will see that changing the structure of the vacuum brane
stress tensor, with the inclusion of higher-rank tensors is crucial
also for type of geometry
described above, as it allows smooths solutions. 
This type of geometry is the higher-codimensional version of that
considered in~\cite{Kaloper:2007ap}. Such type of
regularization was studied in~\cite{deRham:2007pz}, in the context of
compact codimension-two brane worlds, in order to obtain
codimension-two effective actions.     
For the sake of generality we  will thus consider in Section~\ref{section:nonvan} some
higher-codimensional generalizations of the backgrounds considered
in~\cite{deRham:2007pz}, 
that will require bulk higher-rank tensors as well as non-vanishing 
bulk cosmological constant or a bulk $\sigma-$model matter
action~\cite{RandjbarDaemi:2004ni}.

\subsection{Adding the $p$-form field}
{}In order to enrich the vacuum structure of our brane world we include
a $p$-form field in the worldvolume of the blown-up brane $\Sigma$. 
We consider the case of a $(n-1)$-form field strength but could
equivalently consider its dual $(D-n)$-form as in the string
landscape~\cite{Bousso:2000xa}. We require its energy-momentum tensor
to have the block-diagonal form  
\bea
T_m{}^n(F) =
\left(
\begin{array}{cc}
T\delta_\mu{}^\nu & 0\\
0 & T' \delta_\alpha{}^\beta
\end{array}\right)
\label{block-diagonal}
\eea  
with $T,\ T'$ constant. In order to achieve that let us use spherical coordinates
\bea
\gamma_{\alpha\beta}~ d\theta^\alpha d\theta^\beta &=&
      {d\theta_0}^2+\sin^2\theta_0~{d\theta_1}^2
+\sin^2\theta_0\sin^2\theta_1~{d\theta_2}^2\ccr&&+\cdots
+\sin^2\theta_0\cdots\sin^2\theta_{n-3}~{d\theta_{n-2}}^2  
\label{(n-1)-sphere}
\eea
to parameterize the $(n-1)$-sphere and let us consider the extended
magnetic monopole field(s) 
\bea
A_{[n-2]\, \pm} =\sqrt{2}\Phi R_{\epsilon}^{n-1}\Big(\pm c+h(\theta_0)\Big)\, E_{[n-2]} 
\label{A-form}
\eea
with $\Phi$ constant, $E_{[n-2]}$ being the
volume form of the equatorial $(n-2)$-sphere, and
with $h'(\theta) = \sin^{n-2} \theta$, and $h(\pi)=-h(0)$; the field strength
\bea
F_{[n-1]} =\sqrt{2} \Phi R_{\epsilon}^{n-1}\, S_{[n-1]}
\label{p-form}
\eea
is thus proportional to the volume form of the unit $(n-1)$-sphere. The field
configurations~(\ref{A-form}) are defined on the north (south)
hemisphere of $S^{n-1}$ and
the integration constant $c$ is fixed by regularity conditions at the
poles~\cite{Nepomechie:1984wu}.  
 From~(\ref{p-form}) one
immediately obtains
\bea
T_m{}^n(F) = M^{D-3}\Phi^2
\left(
\begin{array}{cc}
-\delta_\mu{}^\nu & 0\\
0 & + \delta_\alpha{}^\beta
\end{array}\right)
\eea  
and
\bea\label{T-m-n}
T_\mu{}^\nu &=& -M^{D-3}\frac{n-2}{R^2_\epsilon}
(\lambda +\varphi^2)\delta_\mu{}^\nu \\
T_\alpha{}^\beta &=& -M^{D-3}\frac{n-2}{R^2_\epsilon}(\lambda -\varphi^2)\delta_\alpha{}^\beta 
\eea
where we have defined $\Phi^2 \equiv  (n-2)
\varphi^2/R_\epsilon^2$. Hence the boundary
conditions~(\ref{match1'}) and~(\ref{match2'}) still hold with the
replacements $\lambda\to\lambda-\varphi^2$ and $\lambda\to\lambda-(2n-3)\varphi^2$ 
respectively. We thus have
\begin{eqnarray}
 &&{2f^2(\epsilon)\over{1-f^2(\epsilon)}}={{  L}\over 2 R_\epsilon}
 \biggl(-\lambda+\varphi^2+n-3\biggr)~,
\label{match1''}\\[2mm]
 &&\frac{D-2}{n-2} {2\kappa f(\epsilon)
 \over{1-f^2(\epsilon)}}={{  L}\over 2 R_\epsilon}
 \biggl( -\lambda+(2n-3)\varphi^2-n+1\biggr)
\label{match2''}
\end{eqnarray}
so that $\lambda$ can be either positive or negative, provided
$\varphi^2$ is large enough. The second condition gives
\bea
R_\epsilon = L~
\sqrt{\frac{D-n}{(n-1)(D-2)}}~\frac{1-f^2(\epsilon)}{4f(\epsilon)} 
 \biggl( -\lambda+(2n-3)\varphi^2-n+1\biggr)
\label{radius}
\eea
that replaced into the first condition yields
\bea
f(\epsilon) = \sqrt{\frac{(n-1)(D-2)}{D-n}}~
\frac{-\lambda+\varphi^2+n-3}{-\lambda+(2n-3)\varphi^2-n+1}
\label{fe}
\eea
that is the equivalent of~(\ref{f}). Note however that now there are
smooth solutions with $f(\epsilon) <1$ regardless of the value of the
brane tension (here parameterized by $\lambda$). For example let us consider
$n=3,~D=7$: in such a case we have $f(\epsilon) =
\sqrt{\frac{5}{2}}~\frac{-\lambda+\varphi^2}{-\lambda+3\varphi^2 -2}$ that
can be smaller than one, provided $\varphi^2$ is large enough.
\FIGURE
{
\epsfig{file=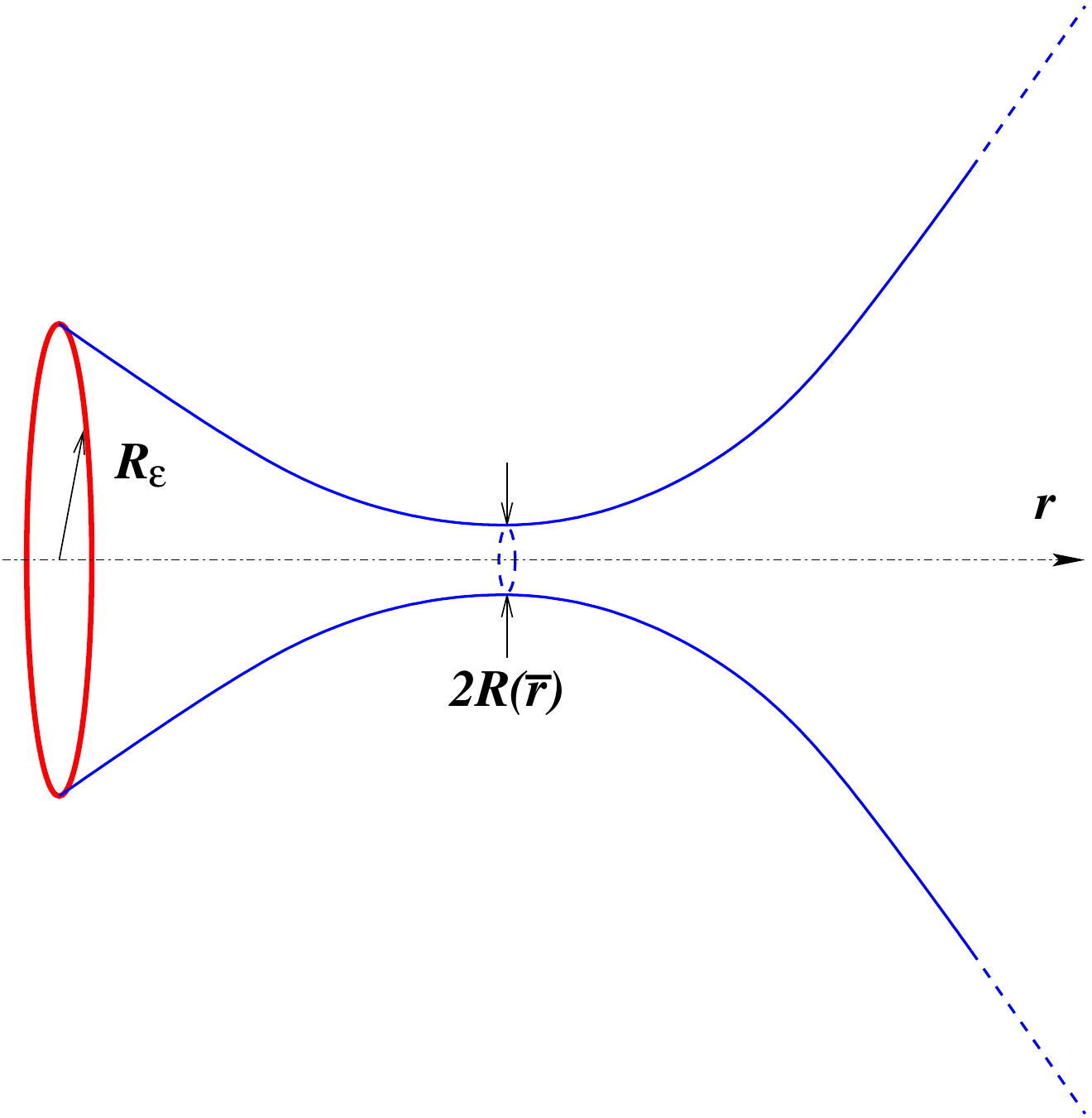,width=6cm}
\caption{Pictorial representation of the extra-dimensional space.}
\label{troath}
}
\noindent The flux
$\varphi^2$ increases the value of the
physical four-dimensional vacuum energy density that can be obtained by
integrating~(\ref{T-m-n}) over the brane profile   
\be
{\cal E}_{4d} = (n-2)S_{(n-1)} ~(\lambda + \varphi^2)~M^{1+n}~R_{\epsilon}^{n-3}
\ee
where $S_{(n-1)}$ is the volume of the unit-radius
$(n-1)-$sphere. Coupling of the form potential to a localized extended object leads
to a quantization condition for the
flux~\cite{Nepomechie:1984wu,Bousso:2000xa,Navarro:2003bf,Nilles:2003km}; we come back to this
point in the next section.

Let us point out a crucial difference between
our setup ($n>2$) and previously considered codimension-two smooth
solution. In the codimension two case~\cite{Kaloper:2007ap} the smooth solution
considered has $A={\rm constant}$ and $B\sim \ln r$ so that the
junction condition coming from the $(\alpha \beta)$ equation of motion~(\ref{AB2})
is trivial (there are no second derivatives in $B$ in such a case) and this can 
only be satisfied if we ``tune'' $\Lambda= \Phi^2$. In our case
the only requirement for the ``flux'' is a lower bound.  Note however
that, once $\lambda$ and $\varphi$ are chosen, the value of the
physical radius is fixed in terms of~(\ref{radius}). 

In the present 4d-Poincar\'e-invariant Ricci-flat setup the inclusion of
bulk fluxes  is problematic because of the no-go theorem~\cite{Maldacena:2000mw}. 
In other words the present solutions avoid such no-go theorem in a
trivial way: no bulk fluxes, and presence of
discontinuities in the derivatives of the warp factors that are
absorbed by localized fluxes and brane tension.

\subsection{Comments}
We comment here on the possible physical scales involved in the
model; we focus on the case $d=4,\ n=3$ for it displays all the details
of these models. 
There is a variety of scenarios that might appear according to
the different values of tension and flux  and it is beyond the scopes
of the present manuscript to give a detailed study of such issues. Let
us however point out a few interesting features. The present
model has codimension larger than two and there is, a priori, no critical value
for the tension. 
\comment{However, one may still try to envisage a 
situation similar to the ``near-critical'' limit
of~\cite{Kaloper:2007ap} where the bulk looks
like a thin cylindrical sliver that ends up on the brane and opens up
non trivially at very large scales. At this limit the see-saw
mechanism~\cite{curvature} explicitly showed up. In our case, casting 
the bulk metric into the form  $ds^2 = V^2(\rho)dx_\mu dx^\mu +d\rho^2 + W^2(\rho)
d\Omega_2^2$ it is obvious that, since the 
internal sphere carries curvature, no Ricci flat solutions are allowed
with $W$ globally constant.} However, 
phenomenological constraints impose that the internal radius of the
brane satisfies $R_\epsilon < (TeV)^{-1}$. Since the scale $L$ will be related
to the crossover scale after which brane gravity turns
higher-dimensional and it must thus be taken to be enormously large,
it is natural to assume that the internal radius
of the brane  be extremely smaller than $L$. Hence, noting that 
a tiny value in the round parenthesis of~(\ref{radius}) would yield
to an inconsistent value for $f(\epsilon)$, equation~(\ref{radius}) implies 
\bea
f(\epsilon)\lsim 1
\label{fine-tuning}
\eea 
that inserted into~(\ref{fe}) yields a fine-tuning relation 
\bea
\varphi^2 \gsim \frac{2-\lx(\sqrt{5/2}-1\rx)\lambda}{3-\sqrt{5/2}}
\label{fine-tuning'}
\eea
that can be satisfied, provided 
\bea
\lambda \leq \lambda_M \equiv 2/(\sqrt{5/2}-1)
\label{tension-contraint}
\eea
that yields a critical value for the brane tension.

Casting the bulk metric into the form  $ds^2 = V^2(\rho)dx_\mu dx^\mu +d\rho^2 + W^2(\rho)
d\Omega_2^2$ one can check that there exists an allowed configuration $\bar
\rho$ at which $W(\rho)$ is critical. For such a value, corresponding
to $\bar r \sim 2 r_*$ the physical
radius $W(\rho) = r\, e^{B(r)}$ assumes its minimum value, after which
the bulk radius asymptotically approaches the flat limit $R(r)\sim r$:
the shape of the extra-dimensional space thus looks like a "throat''
ending on the brane, like depicted in Figure~\ref{troath} and this
scenario is somewhat similar to the ``near-critical'' limit
of~\cite{Kaloper:2007ap} where the bulk looks
like a thin cylindrical sliver that ends up on the brane and opens up
non trivially at very large scales. Then gravity
on the brane should behave $4d$ at distances shorter than $L_C =
\frac{M_4^2}{M_5^3} = \frac{M^{1+n}
  R^{n-1}}{\widehat M^{2+n} R^{n-1}} =L$, then an intermediate $5d$
behavior should take over at distances $\gsim L$ , till the bulk
finally opens up at a scale $L'$ related to $\bar r$, and
brane gravity behaves seven-dimensionally, provided the scale at
which sources on the brane feel the whole seven-dimensional bulk is
larger than $L $. In other words the bulk scale $\bar r$ must be seen
as a very large scale from the point of view of an observer on the 
brane.\footnote{One may worry about the
fact that parallel directions are necessarily warped and it may happen
that a RS-like localization~\cite{Randall:1999vf} takes place at those
scales. More precisely one might expect an interplay of
  effects between induced gravity and RS localization,
  such as the one described in \cite{Kiritsis:2002ca}. However it is
easy to see that ${\rm sign}(V'(\bar \rho)) = {\rm sign}(
Q(\epsilon))>0$ so that, in the near-brane limit the $5d$ ``bulk''
behaves as a brane-to-boundary chunk of AdS and the five-dimensional length
is thus infinite. It is thus natural to expect that no RS localization
takes place.} However regardless of the specific details of the
crossover physics we see that brane tension must be bounded from above
at least for this class of smooth solutions. Also, as mentioned above,
the form potential may be coupled to a charged particle $eM\int_W A$,
where $W$ is the particle worldline (for $n=4$ it would be a string
worldsheet and so on). When $W$ wraps the horizon of the two-sphere,
single-valuedness of the amplitude leads to a quantization condition
for the flux~\cite{Nepomechie:1984wu} $eM\int_{S_2} F =2\pi k$, that yields 
\bea
\varphi R_\epsilon = \frac{k}{2\sqrt2 eM}~,\quad k\in {\mathbb Z}~. 
\eea
Hence, the above fine-tuning relation~(\ref{fine-tuning'}) can be
attained only by the portion of tension that is quantized
accordingly, and the excess of tension $\delta \lambda \sim
\frac1{eM}$ seems to either gravitate or blow the internal radius 
to an unacceptably large value. 
Notice also that flux conservation due to Bianchi identity sets the
conservation of $\varphi R_\epsilon $, similarly to what discussed
in~\cite{Garriga:2004tq} for the finite-volume rugby-ball model. 
In the present setup, unlike what happens in~\cite{Garriga:2004tq}, the flux is not fixed
in terms of bulk parameters and this would, {\em a priori}, allow an
eventual phase transition that locally changes the value of the brane
tension. However, since the internal radius of the brane must locally change
in order to ``absorb'' the different tension and keep the
four-dimensional part of the brane flat, this would lead to a
scenario where different four-dimensional domains (characterized by
different values of tension) have different Planck masses.     

Another issue concerns the stability of such solutions. Although such important
point would require a detailed investigation let us here mention
a few related results obtained in the past in similar models. In the absence of
localized fluxes, instabilities were indeed
found in the models discussed in~\cite{Kakushadze:2001bd,Charmousis:2004zd}. However,
in~\cite{Charmousis:2004zd} it was also shown that localized induced
stress tensors of the form~(\ref{block-diagonal}) do indeed lead to a
stabilization and such effect is quite likely to take place in the
present solutions as well.       

To conclude this section, let us briefly mention that, for generic values of parameters,  
it seems plausible that the higher codimension
resolved brane solutions discussed here behave more like the
``subcritical'' codimension-two cases~\cite{Kaloper:2007ap}, and the crossover scale from
$4d$ gravity and $(4+n)$-gravity is expected to be given by $r_C^n\sim
\frac{M_4^2}{\widehat M ^{2+n}}$.

\section{Non-vanishing bulk cosmological constant}
\label{section:nonvan}
{} In this section we consider some finite-volume counterparts of the
solutions found in the previous section.\footnote{
Higher-codimensional brane
solutions with bulk higher-rank tensors were considered
in~\cite{Navarro:2003bf} where the regularization consisted in blowing
up $n-2$ directions of the brane, hence reducing its codimension from
$n$ to 2. Moreover, in~\cite{Navarro:2003bf}, the ``dual'' form
$\tilde H_{[d]}$ was considered, instead of $H_{[n]}$.   
} What follows is to be
understood as higher-codimension generalizations of the smooth
codimension-two solution described in~\cite{Peloso:2006cq} of which
we also use the conventions. The bulk
part of the action~(\ref{action}) now gets generalized to
\bea
\label{action:finite}
S_{\rm bulk} = \widehat{M}^{D-2} \int d^D x\sqrt{-\widehat g}\Biggl[ 
\widehat {\cal  R} -\widehat \Lambda -\frac1{2(p+1)!}\widehat H_{[p+1]}^2\Biggr] 
\eea   
whereas the brane part remains the same as before. We seek for a
solution with spherically symmetry in the extra $n$-dimensional space and 
Poincar\'e invariance in the $d$ parallel directions
\bea\label{unW}
ds^2 = \eta_{\mu\nu} dx^\mu dx^\nu + R^2\Big(d\theta^2 +\cos^2\theta
d\Omega_{(n-1)}^2
\Big)\,, \quad -\frac{\pi}2 < \theta < \frac{\pi}2
\eea   
and $d\Omega_{(n-1)}^2$ is the line elements of the $(n-1)$-sphere,
explicitly written in~(\ref{(n-1)-sphere}). We locate the smooth
brane at a certain value of the azimuthal angle $\bar\theta$. In fact
in general we might have more branes localized at different
angles. For simplicity we will assume ${\mathbb Z}_2$
symmetry along $\theta$ and a pair of identical branes located at $\pm
\bar\theta$: the symmetry allows us to concentrate only on the
northern hemisphere $\theta >0$. Again, we start considering the case
where $p=n-1$. Similarly to~\cite{Peloso:2006cq} we assume to have an
``inside bulk'' $\bar\theta<\theta<\pi/2$ and an ``outside bulk''
$0<\theta< \bar\theta$ with
different radii, $R_i=R\beta $ and $R_o=R$, and different cosmological constants, 
$\Lambda_i$ and $\Lambda_o$ respectively and we take the magnetic
monopole ansatz for the $n$-form field strength, with
\bea
\widehat H_{[n]} =\left\{ 
\begin{array}{ll}
\widehat Q_i\, (\beta R)^n\ S_{[n]}\,, &\quad {\rm inner\ bulk}\\[2mm]
\widehat Q_o\, R^n\ W_{[n]}\,, &\quad {\rm outer\ bulk} 
\end{array}\right.
\eea 
with $S_{[n]}$ and $W_{[n]}$ respectively being the volume forms of the unit-radius
$n$-sphere and of the unit-radius wedged $n$-sphere whose line element
is given by $d\theta^2 +\beta^2 \cos^2\theta d\Omega_{(n-1)}^2$. 
The bulk equations of motion fix the value of the cosmological
constants and magnetic fields in terms of the radii
\bea
R_a^{-2} = \frac{\widehat \Lambda_a}{(n-1)^2} =\frac{\widehat
  Q_a^2}{2(n-1)}\,,\quad a=i,o~. 
\label{bulk:EoM}
\eea
It is easy to see, that using~(\ref{bulk:EoM}), and redefining
coordinates as
\bea
\theta(l) &=& \bar \theta
-\left(\bar\theta-\frac{l}{R}\right)\Big[\Theta(\bar\theta R
  -l)+\beta^{-1}\Theta(l-\bar\theta R)
  \Big] \\
z^\alpha &=& \beta R \theta^\alpha\,, \quad \alpha=1,\dots,n-1 
\eea 
where $\Theta$'s are Heaviside's step functions, the volume forms
$(\beta R)^n\, S_{[n]}$ and $R^n\, W_{[n]}$ are both given by 
\bea
V_{[n]} = (\cos\theta(l))^{n-1}\ dl\wedge \Omega_{[n-1]}(z)\,, \qquad
ds^2 =dl^2 +\cos^2\theta(l) d\Omega_{(n-1)}^2(z) 
\label{2metric}
\eea
and
\bea
\widehat H_{[n]} = \sqrt{2(n-1)}\, \theta'(l)\ V_{[n]} 
\eea
and $\theta'(l)$ is discontinuous at the location of the brane $\bar
l= \bar \theta R$. The integration by parts associated to the 
equation of motion for the potential form field
\bea
\widehat \omega_{[n-1]} = \Big
         [\pm c+f(\theta(l))\Big]\sqrt{2(n-1)}\,\Omega_{[n-1]}(z)\,,\quad
         f'(\theta) = \cos^{n-1}\theta    
\label{O_n-1}
\eea
whose field strength is $\widehat H$, will
thus give rise to a jump condition at the location of the brane
\bea
\delta_\omega S(H) &\supseteq & -\frac{{\widehat M}^{D-2}}{(n-1)!}\int_{M}d^Dx
\sqrt{-\widehat g}\, \nabla_{M_0}\delta \widehat\omega_{M_1\cdots
  M_{n-1}} \widehat H^{M_0\cdots M_{n-1}}\noN\\
 &=& \frac{{\widehat M}^{D-2}}{(n-1)!}\int_{\partial M}d^{D-1}x
\sqrt{- g}\, \delta \omega^{m_1\cdots
  m_{n-1}} \Big\la \widehat H_{l\,m_1\cdots m_{n-1}}\Big\ra_\pm
\eea
where $\omega_{m_1\cdots m_{n-1}}=\widehat\omega_{m_1\cdots
  m_{n-1}}(\bar\theta)$, and comes from the "-'' branch
of~(\ref{O_n-1}) as the brane sits inside the northern hemisphere. 
 We thus need to ameliorate the  $(n-1)$-form field localized on the
 brane with the inclusion of a
 coupling to $\omega_{[n-1]}$, namely 
\bea
\tilde F_{[n-1]} = F_{[n-1]}+eM \omega_{[n-1]}
\eea   
and
\bea
\delta_\omega S(\tilde F) &=& -\frac{{M}^{D-2}}{(n-1)!}\int_{\Sigma}d^{D-1}x
\sqrt{- g}\, \delta \omega^{m_1\cdots
  m_{n-1}} e\, \tilde F_{m_1\cdots m_{n-1}}~,
\eea
so that 
\bea
\Big\la \widehat H_{l\,m_1\cdots m_{n-1}}\Big\ra_\pm
=L\, M e\, \tilde F_{m_1\cdots m_{n-1}} 
\eea
is the jump condition for the form field, with 
\bea
\Big\la \widehat H_{l\,\a_1\cdots \a_{n-1}}\Big\ra_\pm
=\sqrt{2(n-1)}\frac{1-\b}{R\b}\cos^{n-1}\bar\theta\, \sqrt{\Omega}\,
\epsilon_{\a_1\cdots \a_{n-1}}~.  
\eea 
For the metric we have~(\ref{israel})
instead, that using~(\ref{2metric}), simply yields the following
non-vanishing components for the extrinsic curvature $K_{\a\b}^\pm =
\pm\frac12\partial_l g_{\a\b}=\mp \theta'(\bar l \pm) \tan\bar\theta\,
g_{\a\b}$~. Then,  
choosing the brane magnetic field to be
\bea
F_{[n-1]} = \Phi\, \cos^{n-1}\bar\theta\, \Omega_{[n-1]}\quad\Rightarrow\quad 
\tilde F_{[n-1]} = \tilde\Phi\, \cos^{n-1}\bar\theta\, \Omega_{[n-1]}
\eea
with $\tilde\Phi = \Phi +eM\sqrt{2(n-1)}(f(\bar\theta)-
  c)(\cos\bar\theta)^{1-n}$, we have the following junction conditions
\bea\label{jc1}
\Lambda-\frac12\tilde\Phi^2 &=& \frac{2(n-2)}{L}
\frac{1-\b}{R\b} \tan \bar\theta
\\ \Lambda+\frac12\tilde\Phi^2 &=& \frac{2(n-1)}{L}
\frac{1-\b}{R\b} \tan \bar\theta\label{jc2}
\\ \label{jc3}
e\tilde\Phi &=& \frac{\sqrt{2(n-1)}}{LM}\frac{1-\b}{R\b}~. 
\eea 
where~(\ref{jc1},\ref{jc2}) are the $(\a\b)$ and $(\mu\nu)$ components of
the junction condition for the metric and~(\ref{jc3}) is the junction
condition for the form field. They can be solved to give
\bea\label{jc1'}
\Lambda &=& \frac{2n-3}{L} \frac{1-\b}{R\b} \tan
\bar\theta = \frac{2n-3}{2}\tilde\Phi^2 \\ \label{jc2'}
\tilde\Phi &=& \frac{2eM}{\sqrt{2(n-1)}}\tan\bar\theta 
\eea
so that a brane of arbitrary tension can be accommodated on such a
setup while maintaining $4d$-Poincar\'e invariance. A few observations
are in order. First, it is easy to see that,  contrarily to what
happens in~\cite{Peloso:2006cq}, for fixed $4d$ vacuum energy density
one cannot take the thin limit $\bar\theta \to \pi/2$. The vacuum
energy density can be simply obtained from the integral of the
l.h.s. of~(\ref{jc2}) over the internal profile of the brane. Up to
irrelevant numerical constants it reads
\bea
T \sim \widehat M^{D-2} R^{n-2} (\b \cos\bar\theta)^{n-2}
\sin\bar\theta (1-\b)~.
\eea 
For $n=2$ one recovers the result
of~\cite{Carroll:2003db,Peloso:2006cq}.  
For $n>2$, holding $T$ fixed, the aforementioned limit is impossible as $\beta$
is bounded from above. In other words $T\to 0$ for $\bar \theta \to
\frac\pi 2$; it is thus difficult to imagine how to extend the
approach of~\cite{deRham:2007pz}  to codimension higher that two, at
least within this class of spherically symmetric
regularizations. Also, coupling of the form fields to extended objects
leads to quantization
conditions~\cite{Nepomechie:1984wu,Navarro:2003bf,Nilles:2003km} for
the fluxes that
in turn yield a quantization condition for the brane tension.  

Let us conclude by mentioning possible extensions of the latter
solutions to the case of negative bulk cosmological constant. It
is obvious that an unwarped solution like~(\ref{unW}) with an internal
AdS is prohibited by Maldacena-Nunez no-go
theorem~\cite{Maldacena:2000mw}. However, at least partial way-outs
seem possible if, for instance, one allows the extra space to be
non-compact. In fact, let us start from 
\bea
ds^2 = R^2\Big(\frac{d\xi^2}{\xi^2} +\xi^2\eta_{\mu\nu}dx^\mu
dx^\nu\Big) +\delta_{ab}dz^a dz^b\,,\quad z^a\cong z^a +2\pi l^a
\eea      
where the $(n-1)-$torus parameterized by $z^a$ is the internal profile of the
brane localized at $\xi_0=1$. Bulk equation of motion in presence of
negative cosmological constant and fluxes yields a similar fine-tuning
condition like the one given in~(\ref{bulk:EoM}). Taking for simplicity a
$Z_2$-symmetry and setting $\xi =1+\epsilon |u|$ it is easy to see
that (with the exception of codimension one, where there is no bulk
flux and reduces to the RS2 model~\cite{Randall:1999vf}) finite transverse volume ($\epsilon
=-1$) implies positive brane tension, $\Lambda >0 $ but negative
localized flux, $\tilde \Phi^2 <0$, whereas 
infinite volume   ($\epsilon =+1$) implies negative brane
tension, $\Lambda <0 $ and positive flux, $\tilde \Phi^2 > 0$.

\acknowledgments{This work was partly supported by the Italian
  MIUR-PRIN contract 20075ATT78. The author would like to thank C. Bogdanos,
  C. Charmousis, C. Germani and A. Iglesias for discussions 
  and G. Tasinato for help and critical
  reading of the manuscript. The author is grateful to the LPT Orsay
  for hospitality while parts of this work were completed.}

\end{document}